\documentclass[twocolumn, nofootinbib]{revtex4}

\usepackage{epsf,amssymb,amsmath,latexsym}
\input{epsf.sty}
\begin{document}
\title{Randall-Sundrum Black Holes at Colliders}

\author{Arabella Schelpe}
\email{C.A.C.Schelpe@damtp.cam.ac.uk}
\affiliation{DAMTP, Centre for Mathematical Sciences, University of Cambridge,\\
Wilberforce road, Cambridge CB3 0WA, England}

\begin{abstract}
In this paper we calculate the evolution of black holes at colliders in the
Randall-Sundrum scenario, taking the effect of accretion from the
quark-gluon plasma into account. We calculate the evolution using both
the canonical and microcanonical ensemble approaches and compare the results
to each other and to the well-known corresponding lifetimes for ADD. We
find that for an initial mass of 10TeV the canonical lifetime is of
the order of $10^{-25}s$, which is of the same order of magnitude as
its ADD counterpart. In the microcanonical approach, like in ADD, the microcanonical deviations
from thermality slow down the evaporation of the black hole, but,
unlike in ADD, the black hole does not completely evaporate; it tends
to a stable final state of 0.16GeV. As far
as we are aware this is the first prediction of a stable black hole remnant
using semiclassical gravity.
\end{abstract}
\maketitle

\section{Introduction}
With the imminent arrival of data from the LHC, much recent interest in black
hole events has been focussed on detailed calculations of their
signatures \cite{web, kocetal, hos, mearan, cavetal, gin}. In this
paper we return to a more basic question - that of the lifetimes of
the black holes. An important question is whether instead of harmlessly
evaporating away the black holes might grow. An
argument used to dispatch these worries is that cosmic rays bombard
white dwarfs in much
higher energy collisions than those at the LHC, and so the white dwarfs' observed
lifetimes can be used to put a very low upper bound on the
probability of a catastrophic collision \cite{gidman, jafetal}. Nevertheless, it is still
interesting to calculate the black holes' lifetimes, as the
requirement that they must not grow too large may constrain the models of
extra-dimensional physics which give rise to them.

Mini black holes have been quite extensively studied in the ADD
scenario \cite{web, kocetal, hos, cavetal, gin, ancgol, cavetal2,
    riz, empetal}. Their lifetimes have been calculated and convincingly
shown to be small even when accretion from the quark-gluon plasma is
taken into account \cite{chaetal}. 

Mini Randall-Sundrum (RS) black holes in colliders on the other hand have been much
less studied. Most of the literature that exists has considered the
warping to be negligble and so the analysis is more or less the same
as that for ADD black holes \cite{mearan, sto}.

In \cite{C+H}, Casadio and Harms considered a different kind of black
hole solution \cite{dadetal}, where the warping gives rise to a tidal term on the
four-dimensional brane. They calculated the evolution without accretion
of this black hole and found an exponential decay with typical
lifetime $>10^9s$. This is a surprisingly long time and led us to wonder
whether when accretion is taken into account the black hole might be
stable or even grow.

Therefore, in this paper we calculate the lifetime of this black hole,
including the effect of accretion from the quark-gluon plasma. We give
two alternative calculations. The first is the standard, canonical,
calculation of the evaporation rate, where the black hole is taken to
be in thermal equilibrium with its radiation. The second is the more
accurate, microcanonical, calculation, where the effect on the black
hole mass of losing radiation is taken into account.

The paper is structured as follows. In section \ref{notation} we set up notation. In section \ref{accretion} we derive the accretion rate
in RS, which is the same for both the canonical and microcanonical
calculations. Then in section \ref{can} we calculate the
canonical lifetime. In section \ref{micro} we look at the
microcanonical black hole evolution, and in
section \ref{conclusion} we conclude.

\section{RS Black Holes}\label{notation}

The black hole solution we consider was discovered by
Dadhich et al \cite{dadetal}. Its induced metric on the brane is given by
\begin{equation}ds^2=-A(r)dt^2+A(r)^{-1}dr^2+r^2d\Omega^2,\label{metric}\end{equation}
where
$A(r)=1+\frac{\alpha}{r}+\frac{\beta}{r^2}$,
$\alpha=-\frac{2Ml_p}{m_p}$ and
$\beta=-q\frac{m_p^2l_p^2}{m_5^2}$.
$q=\frac{M}{m_{ew}}\left(\frac{m_p}{m_{ew}}\right)^{\alpha}$ is the
tidal term, where $\alpha$ must be $\geq0$ for black holes of mass
$\sim m_{ew}$ to exist.\footnote{Here, we use $m_p$, $l_p$ to denote
  the 4d Planck mass and Planck length respectively, $m_5$ to denote
  the 5d Planck mass and $m_{ew}$ the electroweak symmetry breaking
  scale.} The horizon is at 
\[r_H=M\frac{l_p}{m_p}\left(1+\sqrt{1+\frac{qm_p^4}{M^2m_5^2}}\right).\]
Some attempts have been made to extend (\ref{metric}) off the brane
\cite{chaetal2}, but as yet no consistent solution has been
found. Hence in this paper we consider only on-brane evaporation, and
hope that, as in ADD\cite{empetal}, the on-brane radiation is dominant.

\section{Accretion}\label{accretion}

We use the same accretion rate as \cite{chaetal}:
\begin{equation}\left.\dot{M}\right|_{\mbox{\tiny accr}}=F\pi r_{\mbox{\tiny eff}}^2\epsilon(t),\label{accrrate}\end{equation}
except that our $r_{\mbox{\tiny eff}}$ is different, as \cite{chaetal} consider a
(4+d)d Schwarzschild black hole. For us, 

\[r_{\mbox{\tiny eff}}^2=\frac{r_c^4}{M}\left(\frac{l_p}{m_p}r_c+\frac{m_p^{2+\alpha}}{m_{ew}^{3+\alpha}}l_p^2\right)^{-1}\!\!,\]
where
\[r_c=\frac{1}{4}\left(6M\frac{l_p}{m_p}+\sqrt{\left(6M\frac{l_p}{m_p}\right)^2+32M\frac{m_p^{2+\alpha}l_p^2}{m_{ew}^{3+\alpha}}}\right).\]
Like \cite{chaetal} we take $\epsilon(t)=517GeV/fm^3$, and the maximum
possible value of $F$, $F=1$, to obtain an upper limit on the
lifetimes of the black holes.

\section{Canonical Evaporation}\label{can}

In 4d the canonical evaporation rate is 
\begin{equation}
\left.\dot{M}\right|_{\mbox{\tiny evap}}=-\sum_s\Gamma_sg_{\mbox{\tiny
    eff}}^sA_3T^4,\label{canrate}\end{equation}
where $T$ is the temperature of the black hole, $A_3$ is its event
horizon area on the brane, $\sum_s$ is a sum over the spins of the
Standard Model
particles into which the black hole evaporates, $g_{\mbox{\tiny eff}}^s$ is the
effective number of degrees of freedom of each spin and $\Gamma_s$ is
the corresponding spin-dependent greybody factor.

Since (\ref{metric}) is the metric of a 4d Reissner-Nordstr\"om black
hole, the temperature is easily calculated:

\[T=\frac{l_p}{2\pi m_pr_H^2}\sqrt{M^2+q\frac{m_p^4}{m_5^2}}.\]

The area of the event horizon on the brane is $4\pi r_H^2$. We use the
$g_{\mbox{\tiny eff}}^s$ values of \cite{chaetal}: $g_{\mbox{\tiny
    eff}}^1=18$ and $g_{\mbox{\tiny eff}}^{\frac{1}{2}}=28$.

The greybody factor, $\Gamma_s$, is composed of two parts: spin-dependent and
geometry-dependent. The spin-dependent part is given by \cite{ancgol}:
$\Gamma_0^{\mbox{\tiny spin}}=1$,
$\Gamma_{\frac{1}{2}}^{\mbox{\tiny spin}}=\frac{2}{3}$ and
$\Gamma_1^{\mbox{\tiny spin}}=\frac{1}{4}$. We approximate the
geometry-dependent part by the geometric optics approximation for 5d
Schwarzschild \cite{hanetal}: $\Gamma_s^{\mbox{\tiny geometry}}=4$. 

Combining (\ref{accrrate}) and (\ref{canrate}) we used Mathematica to
calculate the canonical evolution of the black hole. A graph of mass
vs. time for an initial
mass of 10TeV, in the limiting case of $\alpha=0$, is given in Figure
1. The lifetime is $\tau=1.6\times10^{-25}s$. This is a similar order of magnitude to the ADD lifetimes given by \cite{chaetal}.
\epsfysize=5 cm
\begin{figure}
\center{
\leavevmode
\epsfbox{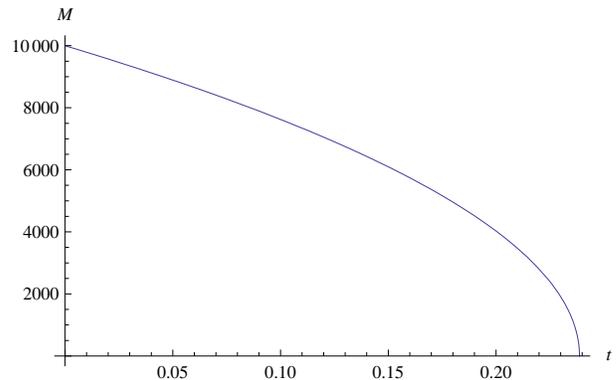}
\caption{Mass (in $GeV$) vs. time (in $GeV^{-1}$) for the canonical evolution of a black hole of
  initial mass 10TeV, in the limiting case of $\alpha=0$.}
}
\end{figure}

It is now interesting to look at how it compares to the
microcanonical evolution.

\section{Microcanonical Evaporation}\label{micro}

The microcanonical luminosity is given by \cite{C+H}:
\begin{equation}
{\cal
  L}_{\mbox{\tiny micro}}(M)=A_3\sum_s\Gamma_s\int_0^{\infty}n(\omega)\omega^3d\omega,\label{microrate}\end{equation}
where 
\[n(\omega)=C\sum_{l=1}^{\lfloor\frac{M}{\omega}\rfloor}\exp(S_E(M-l\omega)-S_E(M)),\]
$C$ is assumed constant, and $S_E(x)$ is the Euclidean action of a
black hole of mass $x$.  

Hence to find ${\cal
  L}_{\mbox{\tiny micro}}(M)$, we firstly need to calculate the Euclidean action of
  (\ref{metric}). This turns out to be
  \[S_E=\frac{m_p}{l_p}\pi\alpha\left(\alpha+\frac{2\beta}{r_H}\right)^{-1}r_H^2.\]
For $M<10TeV$, it can be approximated by 
\begin{equation}S_E=\frac{\pi
    l_p^2m_p}{m_5m_{ew}^{\frac{1}{2}}l_p}M^{\frac{3}{2}}\equiv\left(\frac{M}{m_{\mbox{\tiny eff}}}\right)^{\frac{3}{2}}\equiv m^{\frac{3}{2}}.\label{smallmaction}\end{equation}
This disagrees with the action \cite{C+H} give for the RS black
hole. Their action is $S_E=m$. We believe this difference arises from their
incorrect use of the standard Reissner-Nordstr\"om Euclidean action (as given for example
in \cite{G+H}). The standard RN Euclidean action includes a
contribution from the electromagnetic action,
\[S_{em}=\frac{1}{16\pi}\int d^5x\sqrt{|g|}\frac{1}{2}A_{\nu}\nabla_{\mu}F^{\mu\nu}+\frac{1}{8\pi}\int
d\Sigma\frac{1}{4}A_{\nu}n_{\mu}F^{\mu\nu}.\]
In the case of Dadhich et al's black hole, this contribution should not be
included as there is only an apparent electric charge in the metric
which is not a source for the electromagnetic field.

Substituting (\ref{smallmaction}) into (\ref{microrate}), we now obtain
\begin{equation}{\cal
    L}_{\mbox{\tiny micro}}=K_0K_{\mbox{\tiny small}}me^{-m^{\frac{3}{2}}}\int_0^m
  e^{x^{\frac{3}{2}}}(m-x)^3dx,\label{smallmlum}\end{equation}

In this equation $K_0$ and $K_{\mbox{\tiny small}}$ have between them absorbed all
the multiplicative
constants. $K_{\mbox{\tiny
    small}}=4\pi\frac{l_p^2m_p^2}{m_5^2m_{ew}}m_{\mbox{\tiny eff}}^5$,
and so can be evaluated, but $K_0$ contains a constant which depends
on the quantum description of the
black hole and so cannot be straightforwardly calculated. This poses something of a problem as we would like to
numerically solve the differential equation \[\frac{dM}{dt}=-{\cal
  L}_{\mbox{\tiny micro}}(M)+\left.\dot{M}\right|_{\mbox{\tiny accr}}\] for $M(t)$. However, in
the large $M$ limit the microcanonical and canonical descriptions coincide, and so we can find $K$ by matching the
microcanonical luminosity to the canonical luminosity in this limit.

\subsection{Microcanonical/Canonical Matching}

The large $M$ limit of the canonical luminosity is easy to calculate
as we made no assumptions about the size of $M$ in deriving
(\ref{canrate}). So we can take the large $M$ limit of (\ref{canrate}) to obtain
\begin{equation}
{\cal L}_{\mbox{\tiny can}}(M)=\frac{139}{2^6.720\pi}\left(
\frac{m_p}{l_p}\right)^2M^{-2}.\label{canratelargem}\end{equation}

The large $M$ limit of the microcanonical luminosity on the other hand
is not so easy, as our derivation is only valid for small mass.
For large $x$,
\[S_E(x)=4\pi\left(\frac{l_p}{m_p}\right)^2\frac{m_p}{l_p}x^2\equiv\left(\frac{x}{m_{\mbox{\tiny
      eff}}^{(2)}}\right)^2.\]
Therefore, for large $M$, (\ref{microrate}) becomes
\[{\cal L}_{\mbox{\tiny micro}}(M)=K_0K_{\mbox{\tiny large}}m^2e^{-m^2}\int_0^m e^{S(x)}(m-x)^3dx,\]
where $m=\frac{M}{m_{\mbox{\tiny eff}}^{(2)}}$, $K_0$ is as in
(\ref{smallmlum}), and $K_{\mbox{\tiny
    large}}=4\pi\left(\frac{2l_p}{m_p}\right)^2m_{\mbox{\tiny
    eff}}^{(2)\,6}$. 
This integral is dominated by wherever $S(x)$ is largest. $S$
increases with $x$ and so we can use the large $x$
approximation of $S(x)\sim x^2$ throughout the integral.

Approximating $\int_0^me^{x^2}dx$ by an asymptotic series we obtain
\[{\cal
  L}_{\mbox{\tiny micro}}(M)=\frac{3}{8}K_0K_{\mbox{\tiny large}}\left(\frac{1}{m^2}+O\left(\frac{1}{m^4}\right)\right).\]

Matching this to equation (\ref{canratelargem}), we finally find:
\begin{equation}K_0=\frac{139}{9\cdot120}\pi^2.\label{k}\end{equation}

Now the integral in (\ref{smallmlum}) can be evaluated. 

\subsection{Evolution}

\begin{figure}
\center{
\leavevmode
\epsfbox{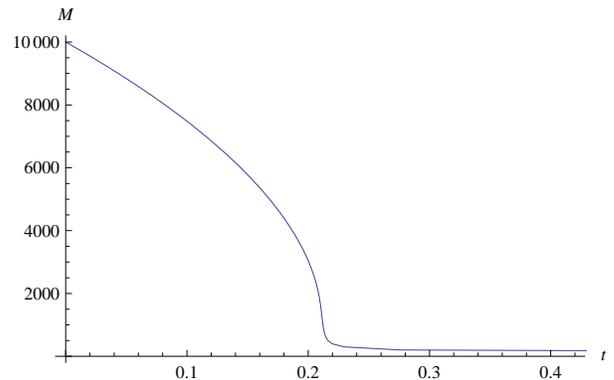}
\caption{Mass (in $GeV$) vs. time (in $GeV^{-1}$) for the microcanonical
  evolution of a black hole of initial mass 10TeV.}
\label{microevol}}
\end{figure}

(\ref{smallmlum}) cannot be evaluated analytically and so we use
Mathematica to solve it numerically. The resulting evolution is shown
in Fig. \ref{microevol}.

\epsfysize=5 cm

The graph demonstrates very clearly the change in behaviour from the
canonical decay shape to the microcanonical decay shape which prolongs
the life of the black hole. The canonical behaviour dominates until
$t\approx0.22GeV^{-1}=1.4\times10^{-25}s$ and the shape of the graph
fits with the lifetime of $1.6\times10^{-25}s$ that we obtained for
the canonical evolution in section \ref{can}. After
$t\approx0.22GeV^{-1}$, the behaviour looks like an exponential decay
asymptoting to zero, as it is in Casadio and Harms' calculation.

However, the decay is in fact slower than exponential. For very small
mass one can calculate an approximate evolution analytically, and find
that $M\sim t^{-\frac{1}{4}}$. It is not clear how to make a
meaningful comparison to Casadio and Harms' result. This is just a
different kind of decay.

A further difference from Casadio and Harms' result is that our black
hole does not completely decay. It evaporates down to a tiny stable
final state of mass 0.16 GeV.

As
far as we are aware, this is the first prediction of a stable final
state using semiclassical gravity. Previous discussion of
stable remnants in the literature have either postulated that instead
of semiclassical gravity being valid below the Planck scale and so the
black hole evaporating completely, the semiclassical description of
black holes breaks down
at the Planck scale and a Planck mass remnant remains \cite{hos,
  kocetal, hosetal}, or they have worked with modifications of
semiclassical gravity which have then predicted Planck mass
size remnants \cite{cavetal2, hosetal2}. 

Since the mass of our stable remnant is less than the Planck mass, our
prediction will only hold if the semiclassical description of black holes is valid below the
Planck scale.

It is not clear how different the experimental signature of such a
small remnant compared to a complete decay would be. It would be
interesting to look at this, but we leave it to future research.
 
\section{Conclusions}\label{conclusion}

We have calculated the evolution of black holes at colliders in the RS
scenario taking accretion into account, which, as far as we are aware,
has not been done before. We calculated the evolution using both the
standard canonical method and the more accurate microcanonical
method. In the canonical case we found that a black hole of initial
mass 10TeV decayed with a lifetime of $\sim 10^{-25}s$. This is of the
same order of magnitude as the lifetimes of similar sized black holes
in ADD. In the microcanonical case we found that as in ADD the
microcanonical deviations from thermality act to prolong the life of
the black hole. We corrected a mistake in a previous calculation of
the lifetime without accretion by Casadio and Harms and found that
when accretion is taken into account the microcanonical
evolution leaves a stable remnant of 0.16GeV. As far as we are aware
this is the first prediction of a black hole final
state using semiclassical gravity. It would be interesting
to study the phenomenological consequences of such a remnant. 

In this calculation, we only considered radiation on the brane, as the black hole
metric has not yet been extended off the brane, so our lifetimes are
upper bounds on the actual lifetimes. It may be that once off-brane
radiation is taken into account the black hole is found to decay
completely in
the microcanonical calculation. Nonetheless, we feel that these
calculations are interesting progress towards understanding the RS
black hole evolution more fully.

\section{Acknowledgements}

I would like to thank David Jennings for collaboration in the early
stages of this project. I would also like to thank Ben Harms and Malcolm Perry for
some technical clarifications.

\end{document}